\pdfoutput=1
\documentclass[aps,twocolumn,superscriptaddress, prl]{revtex4-2}
\usepackage{amsmath}
\usepackage{amssymb}
\usepackage{graphicx}
\usepackage{epstopdf}
\usepackage[colorlinks=true]{hyperref}
\usepackage{physics}
\usepackage{mathrsfs}
\usepackage{comment}
\newcommand{\sectionn}[1]{{\textit{#1}}}
\renewcommand\({\begin{equation}}	
\renewcommand\){\end{equation}}
\renewcommand\[{\begin{eqnarray}}	
\renewcommand\]{\end{eqnarray}}

\DeclareFontFamily{OT1}{pzc}{}
\DeclareFontShape{OT1}{pzc}{m}{it}{ <-> s*[1.26] pzcmi7t }{}
\DeclareMathAlphabet{\mathpzc}{OT1}{pzc}{m}{it}

\newcommand{\al}[1]{\begin{aligned}#1\end{aligned}}
\usepackage[dvipsnames]{xcolor}

\definecolor{mediumtealblue}{rgb}{0.0, 0.33, 0.71}

\begin{document}

\title{Emergent ac Effect  in Nonreciprocal Coupled   Condensates}

\author{Ji Zou}
\affiliation{Physics Department, King Fahd University of Petroleum and Minerals, 31261, Dhahran, Saudi Arabia}
\affiliation{Quantum Center, KFUPM, Dhahran, Saudi Arabia}
\author{Valerii K. Kozin}
\affiliation{Department of Physics, University of Basel, Klingelbergstrasse 82, 4056 Basel, Switzerland}
\author{Daniel Loss}
\affiliation{Physics Department, King Fahd University of Petroleum and Minerals, 31261, Dhahran, Saudi Arabia}
\affiliation{Quantum Center, KFUPM, Dhahran, Saudi Arabia}
\affiliation{Department of Physics, University of Basel, Klingelbergstrasse 82, 4056 Basel, Switzerland}
\affiliation{RDIA Chair in Quantum Computing}
\author{Jelena Klinovaja}
\affiliation{Department of Physics, University of Basel, Klingelbergstrasse 82, 4056 Basel, Switzerland}

\begin{abstract}
We report an emergent ac Josephson-like effect arising without external bias, driven by the interplay between nonreciprocity and  nonlinearity in coupled condensates. Using a minimal model of three mutually nonreciprocally coupled condensates, we uncover a rich landscape of dynamical phases governed by generalized Josephson equations. This goes beyond the Kuramoto framework owing to inherent nonreciprocity and dynamically evolving effective couplings, leading to static and dynamical ferromagnetic and (anti)vortex states with nontrivial phase winding. Most strikingly, we identify an ac phase characterized by the emergence of two distinct frequencies, which spontaneously break the time-translation symmetry: one  associated with the precession of the global U(1) Goldstone mode and the other  with a stabilized limit cycle in a five-dimensional phase space. This phase features bias-free autonomous oscillatory currents beyond conventional Josephson dynamics. We further examine how instabilities develop in the ferromagnetic and vortex states, and how they drive transitions into the ac regime. Interestingly, the transition is hysteretic: phases with different winding numbers destabilize under distinct conditions, reflecting their inherently different nonlinear structures. Our work  lays the foundation for exploring nonreciprocity-driven novel dynamical phases in a broad class of condensate platforms.
\end{abstract}

\date{\today}
\maketitle

Interacting  condensates exhibit rich dynamical phenomena. A hallmark example is the Josephson effect~\cite{josephson1962possible}, where a static phase difference induces a dc current without external bias, while a frequency mismatch, typically driven by a bias, leads to an ac current. These phenomena have far-reaching impact across physics, from condensed matter and atomic systems to quantum information science. The study of coupled condensate dynamics now spans a diverse array of platforms, including ultracold atomic gases~\cite{anglin2002bose},  exciton–polariton condensates~\cite{deng2010exciton}, photonic systems~\cite{bloch2022non}, and magnon condensates~\cite{demokritov2006bose,bozhko2016supercurrent,kreil2021experimental}. 

In realistic settings, condensates inevitably interact with their environment, leading to decay. While dissipation is traditionally regarded as detrimental~\cite{lu2016geometrical,keeling2008spontaneous,szymanska2006nonequilibrium}, recent studies have revealed that dissipative couplings (or collective decay), either intrinsic or engineered, can enable nonreciprocal interactions. Such nonreciprocity has emerged as  a powerful resource for controlling transport and nonequilibrium phenomena, giving rise to diode-like behavior~\cite{zou2024prl,yuan2023unidirectional}, unidirectional amplification~\cite{metelmann2015nonreciprocal}, enhanced quantum entanglement~\cite{reiter2016scalable,zou2022prb,zou2024spatially,driessen2025robust}, and nonreciprocal phase transitions~\cite{fruchart2021non}. Yet,  the role of nonreciprocity in the dynamics of coupled condensate systems remains largely unexplored. This raises an interesting question: What new phenomena emerge when nonreciprocal interactions are introduced into phase-coherent interacting condensates with intrinsic nonlinearity?

\begin{figure}
	\centering\includegraphics[width=0.92\linewidth]{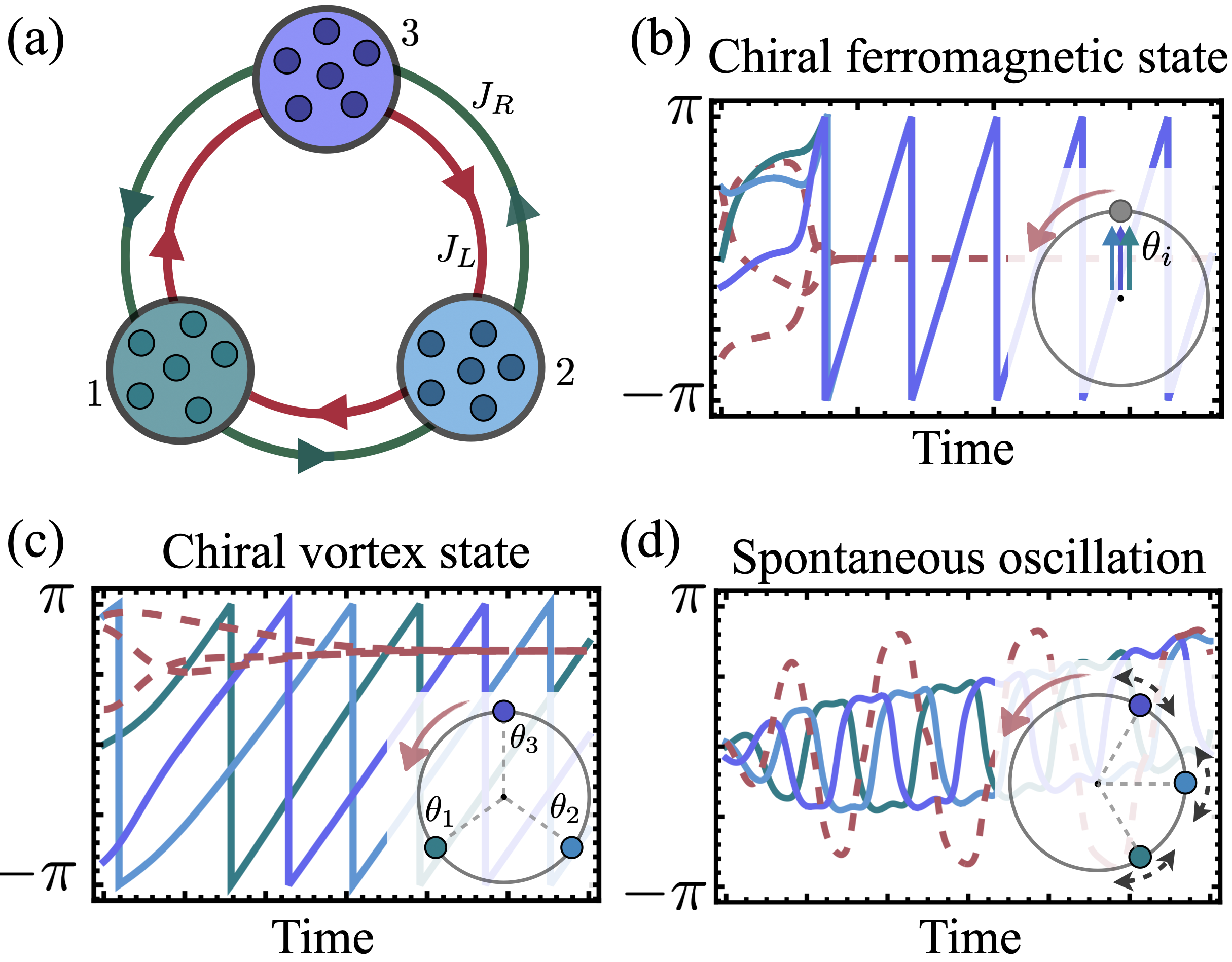}
 \caption{(a) Schematic of three mutually nonreciprocally coupled condensates, with asymmetric couplings $J_L$ and $J_R$.
(b) Chiral ferromagnetic phase, where all condensates share the same phase and undergo collective precession at a constant frequency. (c) Chiral vortex phase with a phase winding of $2\pi$; relative phases are locked while the global phase rotates uniformly. (d) Spontaneous ac phase, featuring two emergent frequencies: one associated with global phase rotation and the other with oscillations of  relative phases. Dashed red: relative phase $\theta_{ij}$; solid lines: individual phases $\theta_i(t)$~\cite{fig1parameter}.   }
  \label{fig1}
\end{figure}

Here, we show that nonreciprocity, together with nonlinearity, gives rise to a novel dynamical phase, featuring the spontaneous emergence of two distinct frequencies and self-sustained oscillatory currents in the absence of any external bias, going beyond the conventional ac Josephson effect and reminiscent of time-crystalline behavior~\cite{Sacha2017, RevModPhys.95.031001, PhysRevLett.114.251603, PhysRevLett.123.210602}. We demonstrate this phenomenon by considering a minimal model of three nonreciprocally coupled condensates, as illustrated in Fig.~\ref{fig1}(a). By deriving the generalized Josephson equations for the system, we reveal how the interplay between nonreciprocal interactions and intrinsic nonlinearity give rise to a rich landscape of dynamical phases. We find both static and dynamical (chiral) ferromagnetic states, where all phases are locked [see Fig.~\ref{fig1}(b)], as well as (anti)vortex states characterized by nonzero phase winding, as shown in Fig.~\ref{fig1}(c). We further investigate how instabilities develop within these phases and trigger transitions into the ac regime [see Fig.~\ref{fig1}(d)]. Interestingly, we find that the instability is path-dependent: distinct winding numbers destabilize under different conditions, revealing how topology shapes the nonlinear structure. We understand the emergence and transition between these phases using insights from bifurcation theory.

\sectionn{Model and generalized Josephson equations}|We consider a model of three coupled condensates, each described by a macroscopic wavefunction $\psi_i = \sqrt{N_i} e^{i\theta_i}$ for $i \in \{1, 2, 3\}$, where $N_i$ and $\theta_i$ denote the particle number and phase, respectively. The system evolves according to~\cite{limit_cycle_note1} $i\hbar \partial_t \psi_i=(\Delta_i-i\gamma)\psi_i +J_L \psi_{i+i} +J_R\psi_{i-1}$.
We will assume no external bias, i.e., $\Delta_i = \Delta_0$ for all $i$, and set $\Delta_0 = 0$ by moving to the rotating frame. 
 The first term accounts for local losses in each condensate with rate $\gamma>0$, while the last two terms capture nonreciprocal coupling between them, with $J_L = \mathcal{J} - iG/2 = J + i(D - G/2)$ and $J_R = \mathcal{J}^* - iG/2 = J - i(D + G/2)$. Here, $\mathcal{J}=J+iD$ denotes the  coherent coupling consisting of symmetric $J$ and antisymmetric $D$ components. An important ingredient in this study is the parameter $G$, which quantifies the strength of the dissipative coupling. Stability of the system requires  $|G|\leq \gamma$ (to avoid net gain).   In different physical platforms, such coupling can arise through various mechanisms. For instance, in magnon condensates, the dissipative coupling naturally emerges via nonlocal  spin transfer across nonmagnetic spacers~\cite{heinrich2003dynamic,nakata2024magnonic,yu2024non}; in photonic systems, it can be engineered through tailored reservoirs or auxiliary lossy modes~\cite{metelmann2015nonreciprocal,PhysRevLett.123.233604,zhen2015spawning,miri2019exceptional}; and in ultracold atom setups, nonreciprocal coupling can be implemented using chiral waveguides or structured optical reservoirs~\cite{gou2020tunable,PhysRevLett.129.070401,reisenbauer2024non}.
 
 We now derive the  generalized Josephson equations for the system.
For notational clarity, we define the unit vector $\vec{n}(\theta_{ij}) \equiv (\cos\theta_{ij}, \sin\theta_{ij})$ associated with the relative angle of the condensates $\theta_{ij} = \theta_i - \theta_j$, and introduce time-dependent effective coupling vectors:
\begin{align}
    \vec{J}_i^{\,L}(t) &= \sqrt{N_{i+1}/N_i} (\Re J_L, \Im J_L), \\
    \vec{J}_i^{\,R}(t) &= \sqrt{N_{i-1}/N_i} (\Re J_R, \Im J_R),
\end{align}
which describe clockwise and counterclockwise particle tunneling, respectively. The generalized Josephson equations then take the compact form: 
\(d\theta_i/dt= -  \vec{J}_i^{\,L}(t) \cdot  \vec{n}(\theta_{i,i+1}) -  \vec{J}^{\,R}_i(t) \cdot  \vec{n}(\theta_{i,i-1}). \label{Eq:theta}  \)
Importantly, the couplings  $\vec{J}_i^{\,L/R}(t)$ depend dynamically on the particle numbers $N_i(t)$, which evolve as:
\( \al{ d N_i/dt= &- 2 N_i  \hat{z}\cdot [ \vec{J}_i^{\,L}(t) \times \vec{n}(\theta_{i,i+1})   
 \\& + \vec{J}^{\,R}_i(t)\times  \vec{n}(\theta_{i,i-1})] -2\gamma N_i +\mathcal{P},  }  \label{Eq:number} \)
where $\mathcal{P}$ denotes a constant pump just used to sustain finite condensates. We choose $\mathcal{P}=\gamma$ without loss of generality. 
These equations exhibit strong nonlinearity with dynamical feedback between particle numbers and phases.
 We note that the system is invariant under a global phase shift $\theta_i \rightarrow \theta_i + \varphi$, reflecting a continuous U(1) symmetry with no preferred phase. As we will show, this symmetry can be spontaneously broken in static phases and dynamically restored in  chiral phases. Moreover, the absence of explicit driving guarantees time-translation invariance, which can likewise be  broken.

We stress that Eq.~\eqref{Eq:theta} goes beyond the  Kuramoto model of synchronization with two important features: the dynamical nature of the coupling strengths and the intrinsic nonreciprocity of the interactions.  The full phase space of this coupled condensate system is  six-dimensional, forming a manifold $\vb{V}=(\theta_i, N_i)\in \mathbb{T}^3\times \mathbb{R}_+^3$. The dynamics on this manifold are inherently nonlinear and nonreciprocal, leading to  quite intricate nonequilibrium behavior.

As a starting point and also to build physical intuition, we begin by considering the steady state case where the particle numbers  are  constant and $N_i=N_0=\text{constant}$ due to the rotation symmetry~[Fig.~\ref{fig1}(a)]. In this case, the model [Eqs.~\eqref{Eq:theta} and~\eqref{Eq:number}] reduces to the nonreciprocal Kuramoto model~\cite{fruchart2021non}. To gain insight, we examine the interaction on a single link. We  write the evolution equations as
\begin{align}
    {d\theta_1}/{dt} &= -J\cos\theta_{21} + (D - G/2)\sin\theta_{21}, \\
    {d\theta_2}/{dt} &= -J\cos\theta_{21} + (D + G/2)\sin\theta_{21}.
\end{align}
We observe that the coherent couplings $J$ and $D$ contribute equally to the evolution of the phases,  modifying the overall rotation rate. In contrast, the dissipative coupling $G$ directly governs the evolution of the relative phase via $d\theta_{21}/dt = G\sin\theta_{21}$. When $G < 0$, this term favors alignment of neighboring phases, effectively acting as a ferromagnetic interaction that leads the three condensates to synchronize into a uniform phase-locked state. In sharp contrast, for $G > 0$, the dissipative coupling favors anti-alignment between neighboring phases. This would  introduce frustration in the three coupled condensates. In this regime, such dissipation-induced frustration  leads to  non-collinear configurations, resulting in a vortex or antivortex state characterized by relative phases of $\pm 2\pi/3$  across the three sites, yielding a total winding of $\pm 2\pi$. Thus, in the nonreciprocal Kuramoto model~\cite{limit_cycle_sm}, the final state  is either a ferromagnetic configuration [Fig.~\ref{fig1}(b)] or an (anti)vortex state [Fig.~\ref{fig1}(c)], dictated solely by the sign of the dissipative coupling $G$. A  detailed analysis, along with numerical results, is provided in SM~\cite{limit_cycle_sm} for the nonreciprocal Kuramoto model. We next examine the stability and evolution of these phases in the full dynamics governed by Eqs.~\eqref{Eq:theta} and \eqref{Eq:number}. 

\begin{figure}
	\centering\includegraphics[width=0.86\linewidth]{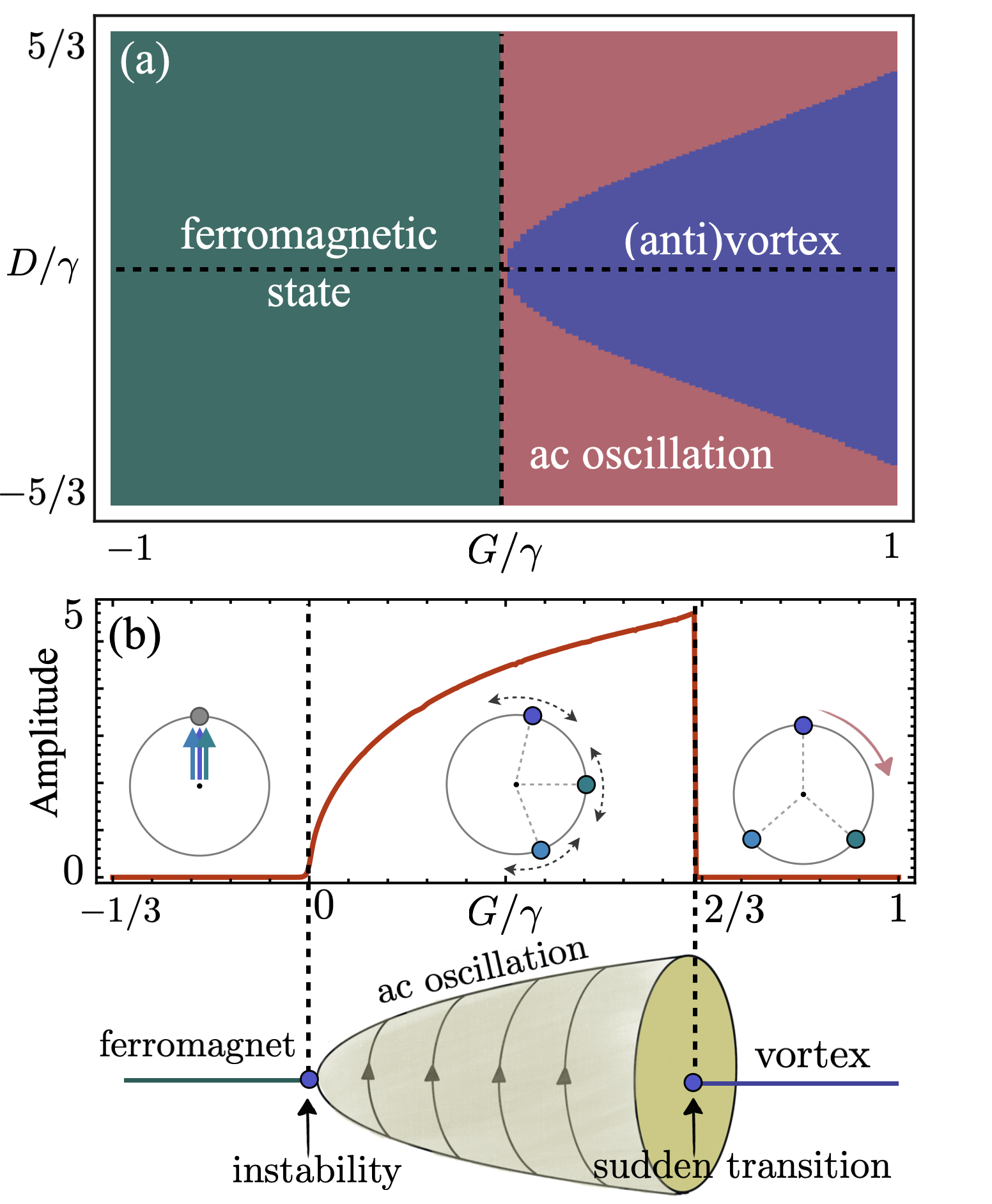}
 \caption{(a) Phase diagram showing the ferromagnetic state and its transition to other phases upon instability for $J=0$. It is obtained by numerically solving the generalized Josephson equations starting from initial conditions near the ferromagnetic fixed point~\cite{Fig2parameter}. (b) Steady-state oscillation amplitude of $\theta_{21}$ as a function of dissipative coupling $G$ for $D/\gamma = 1$ and $J = 0$ obtained~\cite{Fig2parameter}. }
  \label{fig2}
\end{figure}

\sectionn{Ferromagnetic phase and its instability}|Here, we focus on the ferromagnetic state and investigate how instabilities arise, driving transitions to other dynamical regimes. Although the nonequilibrium nature of the system precludes a  free energy description, distinct phases and their transitions can be identified by analyzing  the  generalized Josephson equations~\eqref{Eq:theta} and \eqref{Eq:number}.  While the full dynamics remain analytically intractable, we assess the stability of the ferromagnetic state by introducing small perturbations $\delta \vb{V}$ around the ferromagnetic steady state $\vb{V}_F$, leading to $\partial_t \delta \vb{V}=\hat{L}_F \delta \vb{V}$. Here, $\hat{L}_F$ is a linear operator acting on the tangent space of the phase-space manifold at $\vb{V}_F$. A key feature of this operator is that it always possesses a zero eigenvalue, corresponding to the Goldstone mode associated with the  broken U(1) symmetry, namely the global phase $\Phi = \sum_i \theta_i / 3$~\cite{centermass}.

To isolate the nontrivial dynamics, we exploit this  symmetry and decompose the phase degrees of freedom into the zero mode $\Phi$ and two independent relative phases, $\theta_{21}$ and $\theta_{32}$. This allows us to reduce the dynamics to a five-dimensional effective phase space $\vb*{\mathcal{V}} = (\theta_{21}, \theta_{32}, N_i) \in \mathbb{T}^2 \times \mathbb{R}_+^3$, which we denote as $\mathcal{M}$. The stability of the ferromagnetic phase is governed by the reduced linear operator (see SM~\cite{limit_cycle_sm}), which has the  structure:
\( \hat{\mathcal{L}}_F =\mqty(  \mathcal{L}_F^{\theta} &  J \mathcal{L}_1 \\ J\mathcal{L}_2  &  \mathcal{L}_F^{N} ). \label{eq:linearF}  \)
 It acts on  $\mathcal{M}$ around the ferromagnetic fixed  point $\vb*{\mathcal{V}}_F = (0, 0, N_i = N_0)$, with $N_0 = \gamma / [2(\gamma + G)]$.  Here, $\mathcal{L}_F^{\theta}$ and $\mathcal{L}_F^N$ act on  $\mathbb{T}^2$ and $\mathbb{R}^3_+$ respectively, while $\mathcal{L}_1$ and $\mathcal{L}_2$ encode coupling between them.
 The ferromagnetic phase remains stable when $\hat{\mathcal{L}}_F$ is Hurwitz (all eigenvalues have negative real parts), ensuring that perturbations are damped over time. Instability emerges when any eigenvalue crosses into the right half-plane, marking a dynamical phase transition. We show that the ferromagnetic phase is stable when~\cite{limit_cycle_sm}
\begin{equation}
    G(G + 4\gamma) < 12J^2. \label{eq:9}
\end{equation}
Unlike the nonreciprocal Kuramoto model~\cite{limit_cycle_sm}, where the ferromagnetic phase is stable only for $G < 0$, we find that coherent coupling $J$ extends its stability into the $G > 0$ regime. This arises from the structure of $\hat{\mathcal{L}}_F$ and reflects the nontrivial interplay between  phase and number dynamics: the number sector, governed by $\mathcal{L}_F^N$, is intrinsically stable~\cite{limit_cycle_sm}, and its coupling through $J$ to the phase sector suppresses emerging instabilities. In this way, number dynamics acts as a dynamical stabilizer, reinforcing the robustness of the ferromagnetic phase beyond its decoupled limit.

We show the phase diagram in Fig.~\ref{fig2}(a) for $J = 0$ (see SM for more cases~\cite{limit_cycle_sm}). Interestingly, the ferromagnetic phase does not immediately transition into a (anti)vortex state. Instead, it first enters a dynamical regime characterized by persistent oscillations on the reduced phase space $\mathcal{M}$, with no equilibrium analogue.  When the dissipative coupling $G$ exceeds a critical threshold, the system eventually settles into an (anti)vortex state.
Figure~\ref{fig2}(b) shows the steady-state oscillation amplitude of $\theta_{21}$ for fixed $D$. For $G < 0$, the system resides in a ferromagnetic state that breaks the global U(1) symmetry. This state is generically dynamical for finite $J$: the Goldstone mode $\Phi$ exhibits uniform rotation at frequency $\Omega = -2J$, dynamically restoring U(1) symmetry on average, a hallmark of the chiral phase in nonreciprocal systems~\cite{fruchart2021non}.  
 However, unlike previous works that rely on stochastic noise to stabilize chiral phases~\cite{fruchart2021non}, the interacting condensate system realizes them purely through its internal dynamical structure, without invoking any other ingredients.  We show such chiral ferromagnetic rotation in Fig.~\ref{fig3}(a)  and also include  an example  in Supplemental Video 1~\cite{limit_cycle_sm}, starting from an initial state near the ferromagnetic fixed point.

As $G$ becomes positive, the system enters a richer dynamical regime: beyond global phase rotation (with  frequency proportional to $J$), the system develops autonomous oscillations on $\mathcal{M}$ at a distinct frequency.  To visualize this, we project the oscillatory trajectory onto the $\mathbb{T}^2$ submanifold  in Fig.~\ref{fig3}(b). These persistent relative-phase oscillations  generate oscillatory particle currents between condensates, accompanied by corresponding modulations in particle numbers. Interestingly, this ac-Josephson-like effect emerges intrinsically from the nonreciprocity and nonlinearity of the system. It is principally distinct from conventional ac effects driven by external bias, which is set to zero in this study.  We show this ac phase, with two emergent frequencies, in Supplemental Video 2~\cite{limit_cycle_sm}.
From the perspective of dynamical systems, the transition from the  ferromagnetic state to persistent ac oscillations is a manifestation of a Poincaré–Andronov–Hopf bifurcation~\cite{chow2012methods} occurring on the five-dimensional manifold $\mathcal{M}$, which originates from a mechanism distinct from that of the chiral phase.  At the Hopf threshold $G_c = 2\left[(3J^2 + \gamma^2)^{1/2} - \gamma\right]$, oscillations of relative angles emerge with amplitude $A$ scaling as $A^2\propto G/2 - \gamma + \sqrt{(G+\gamma)^2 - 9J^2}$ (for the coefficient of proportionality see SM~\cite{limit_cycle_sm}).  In Fig.~\ref{fig3}(b), we observe a clear increase in oscillation amplitude with increasing $G$; its dependence on $J$ is also displayed in Fig.~\ref{fig3}(a). 
Meanwhile,  the oscillation frequency is shown to be $\omega = \sqrt{3}D + \mathcal{O}(G - G_c)$ [see inset of Fig.~\ref{fig3}(b)]. The ac phase therefore exhibits spontaneous breaking of time-translation symmetry through the emergence of two independent frequencies, reminiscent of time-crystalline behavior~\cite{Sacha2017}: one originates from the coherent rotation of the U(1) zero mode (with frequency $\Omega$) and the other from stable limit cycles on $\mathcal{M}$ (with frequency $\omega$).

The ac oscillation phase remains stable until the limit-cycle amplitude exceeds a critical threshold, as shown in Fig.~\ref{fig2}(b). Beyond this point, the trajectory on $\mathcal{M}$ escapes the oscillatory attractor and transitions into a topologically distinct fixed point  corresponding to the (anti)vortex phase (see Supplemental Video 3~\cite{limit_cycle_sm}). 
This sudden change is a manifestation of  a global bifurcation~\cite{chow2012methods},  fundamentally different from the Hopf-type transition [see Fig.~\ref{fig2}(b)] that initiates the ac phase.

\begin{figure}
	\centering\includegraphics[width=0.98\linewidth]{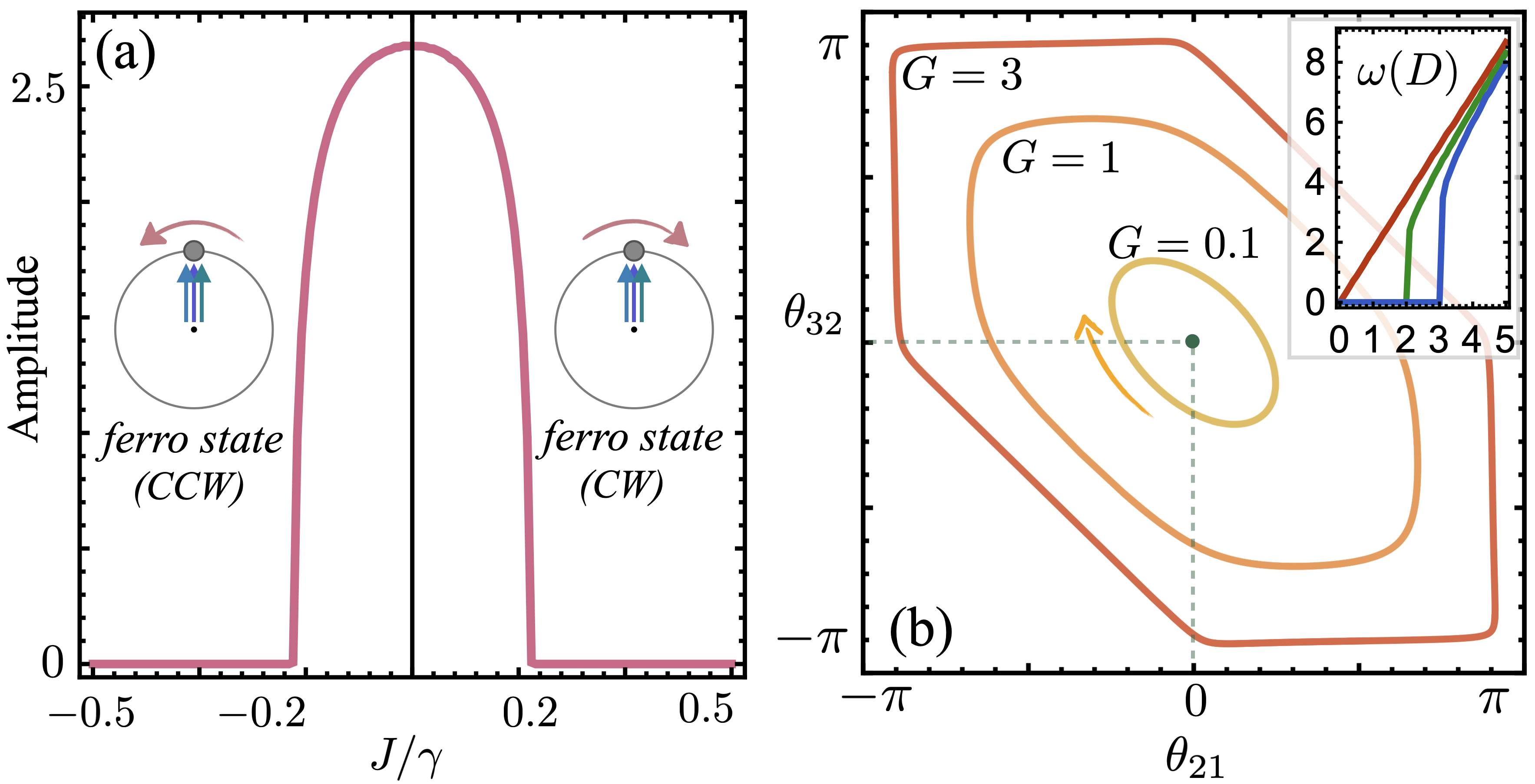}
 \caption{(a) Steady-state oscillation amplitude of $\theta_{21}$ as a function of coupling $J$ for $D/\gamma = 1.2$ and $G/\gamma = 0.1$. (b) Projected trajectories of persistent oscillations on the reduced phase space $\mathcal{M}$ at $J=0$, shown on the $\mathbb{T}^2$ submanifold of relative phases, for varying values of $G$ (with $\gamma = 3$, $D = 5$).
        Inset: oscillation frequency as a function of $D$, for $G = 0$ (red), $G = 1$ (green), and $G = 3$ (blue), with $\gamma = 3$ fixed~\cite{Fig2parameter}. }
  \label{fig3}
\end{figure}

\begin{figure}[!t]
	\centering\includegraphics[width=0.95\linewidth]{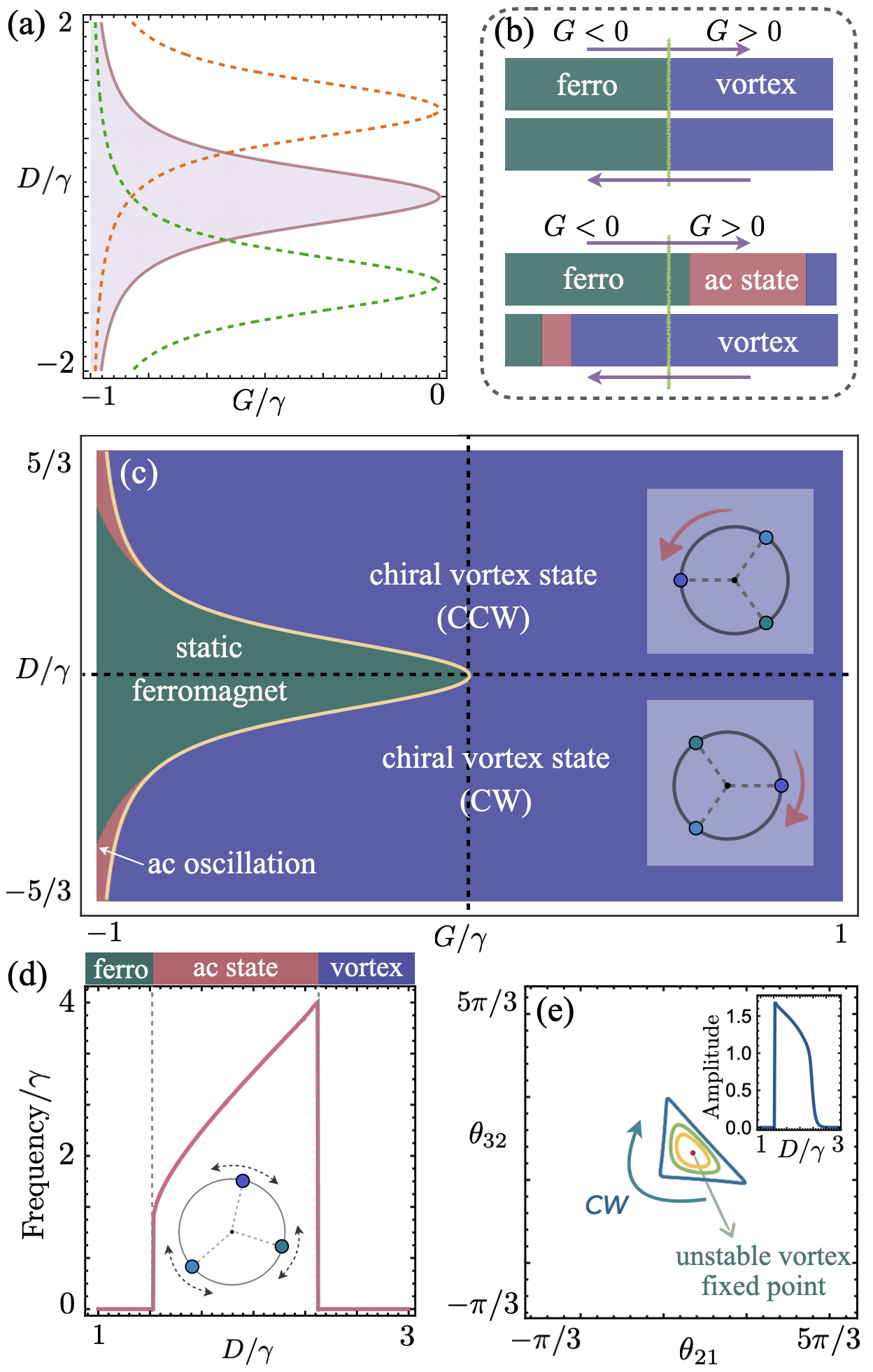}
 \caption{(a) Instability boundary of the vortex phase for $J/\gamma = 0, -\sqrt{3}, \sqrt{3}$ (red, orange, and green curves). The shaded region is the parameter regime where the vortex state is unstable for $J=0$.
 (b)Upper panel: In the nonreciprocal Kuramoto model~\cite{limit_cycle_sm}, the ferromagnetic (green) and vortex (purple) states lose stability simultaneously at $G=0$ and switch directly into each other. Lower panel: In the full dynamics, the ferromagnetic state destabilizes at positive $G$ into an intermediate ac state (red) before reaching the vortex phase, while the vortex state becomes unstable at negative $G$. (c) Phase diagram showing the vortex phase and its transition to other phases for $J=0$.  The yellow curve shows the analytically obtained instability boundary given by Eq.~\eqref{vortexcondition}. 
 (d) Frequency $\omega$ of the steady state as a function of $D$ for $J=0$ and $G/\gamma
 =-0.98$. (e) Limit‑cycle trajectories projected onto the relative‑phase torus $\mathbb{T}^2$ for $J=0$ and $G/\gamma=-0.98$. From inner to outer trajectories, we set $D/\gamma=2.7$, $2.6$, $1.5$. Inset: oscillation amplitude versus $D$.  Plots (c)-(e) are obtained from numerical solutions of the generalized Josephson equations, starting near the vortex fixed point~\cite{fig4parameter}. }
  \label{fig4}
\end{figure}

\sectionn{Instability of vortex state}|Having reached the vortex phase, we now turn to its stability. Interestingly, starting from the vortex state reveals a distinct phase diagram and instability condition compared to Fig.~\ref{fig2}(a). To analyze this, we introduce small perturbations around the vortex fixed point, $\vb*{\mathcal{V}}_V\! =\! (2\pi/3, 2\pi/3, \tilde{N}_0)$, with $\tilde{N}_0 = \gamma / (2\gamma - G)$. This leads to  a linear operator $\hat{\mathcal{L}}_V$~\cite{limit_cycle_sm}, which governs fluctuations on   $\mathcal{M}$ near $\vb*{\mathcal{V}}_V$. Importantly, the structure of $\hat{\mathcal{L}}_V$ is qualitatively distinct  from that of $\hat{\mathcal{L}}_F$ for the ferromagnetic phase. The vortex phase is found to be destabilized when
\begin{equation} 
G(G + 4\gamma)^2 + 72(D + J/\sqrt{3})^2 (G + \gamma) < 0. \label{vortexcondition}
\end{equation} 
We show this instability boundary in Fig.~\ref{fig4}(a) for different exchange coupling $J$.
We stress that this  condition differs markedly from that of the ferromagnetic state~\eqref{eq:9}, since the analysis is carried out around different fixed points with different topology. It is clear that in certain parameter region, both vortex and ferromagnetic phases are stable~\cite{limit_cycle_sm}, which gives rise to the hysteresis  discussed below.
 This is in sharp contrast to the nonreciprocal Kuramoto model~\cite{limit_cycle_sm}, where the transition is symmetric and the instability is always triggered at $G=0$ from both directions as shown in the upper panel of Fig.~\ref{fig4}(b). The different nonlinearity around different phases leads to  hysteretic behavior: the onset of instability depends on the direction of the transition, shown in the lower panel of  Fig.~\ref{fig4}(b). In other words, both the ferromagnetic and (anti)vortex states can coexist as stable fixed points for the same parameters~\cite{limit_cycle_sm}, and the selected state depends on the history (initial condition or sweep direction), giving rise to hysteretic phenomena.  

We  present the phase diagram in Fig.~\ref{fig4}(c) for $J = 0$, obtained by solving the dynamics  numerically by starting from the vortex phase in contrast to Fig.~\ref{fig2}(a) where we start from  a ferromagnetic state. The yellow curve in Fig.~\ref{fig4}(c)  shows the analytically obtained instability boundary~\eqref{vortexcondition}, which agrees well with the numerics. We note that the vortex state persists within a finite window of negative dissipative coupling, $G<0$. This stability, absent in the nonreciprocal Kuramoto model~\cite{limit_cycle_sm}, stems from dynamical feedback: the always‑stable number sector ($\mathbb{R}_+^{3}$) couples to the phase sector ($\mathbb{T}^2$) via the antisymmetric exchange $D$, effectively renormalizing the phase dynamics and shifting the vortex instability to more negative $G$. A larger $D$ strengthens this feedback, further widening the stability window as observed in Fig.~\ref{fig4}(c).  This vortex phase is characterized by a rotating Goldstone mode $\Phi$ with frequency $\Omega=J+\sqrt{3}D$. As a result, time-translation symmetry is spontaneously broken, while the global U(1) symmetry is dynamically restored on average. We show the stabilization of this chiral vortex state in Supplemental Video 4~\cite{limit_cycle_sm}.

The emergence of the ac phase hinges  on the strength of $D$.  For small $|D|$, coupling between phase ($\mathbb{T}^{2}$) and density ($\mathbb{R}_{+}^{3}$) dynamics is insufficient to stabilize a limit cycle, causing a direct collapse from the vortex into a static ferromagnetic state, where global U(1) symmetry is broken but time-translation symmetry is restored. 
Once $|D|$ exceeds a critical threshold, a Poincaré–Andronov–Hopf bifurcation gives rise to a stable limit cycle in the five-dimensional reduced phase space $\mathcal{M}$, thereby generating an ac phase that intervenes between the vortex and ferromagnetic regions. 

Importantly, this regime differs from the ac phase nucleated by instability of the ferromagnetic state: here, the trajectory explores a distinct region of phase space $\mathcal{M}$, as revealed by projections onto the $\mathbb{T}^{2}$ submanifold [Fig.~\ref{fig4}(e)]. This ac phase exhibits robustness against small variations in the coupling strengths~\cite{limit_cycle_sm}, indicating that it does not rely on  fine-tuning and should be accessible in realistic settings.
This ac phase again is  distinguished by the appearance of two  emergent frequencies (see Supplemental Video 5~\cite{limit_cycle_sm}),  spontaneously breaking time-translation symmetry. In addition to the global U(1) Goldstone mode, a second frequency arises from stable limit cycles in the phase space $\mathcal{M}$. As shown in Fig.~\ref{fig4}(d), this limit cycle frequency grows with $D$ but exhibits nonlinear dependence on other coupling parameters. As one decreases $|D|$, the amplitude of the trajectory grows [see Fig.~\ref{fig4}(e) inset], until a global bifurcation abruptly ejects the system into the ferromagnetic fixed point (see Supplemental Video 6~\cite{limit_cycle_sm}), marking a sharp boundary for the ac phase. 

Our results reveal that the interplay between nonreciprocity and nonlinearity can serve as a robust mechanism to engineer novel spontaneous dynamical phenomena, which may be explored in diverse experimental platforms.

\begin{acknowledgments}
\textit{Acknowledgments.}  We thank Even Thingstad and Kouki Nakata for insightful discussions. This work was
supported by the Georg H. Endress Foundation, by the Swiss National Science Foundation, NCCR SPIN (grant no. 51NF40-180604), and by the Ibn Sina Global Scholarship. DL acknowledges the Deanship of Research at
KFUPM and the Quantum Center for the support received under Grant no. CUP25102 and no. INQC2500, respectively.
\end{acknowledgments}


\begin{thebibliography}{38}
\bibitem [{fig()}]{fig1parameter}%
  \BibitemOpen
  \href@noop {} {}\bibinfo {note} {In (b), we use $G=-1/2$, $J=-1$, $D=1$ and
  initial conditions $\theta_1(0)=0, \theta_2(0)=\pi/2, \theta_3(0)=-\pi/5$; In
  (c), we use $G=-1$, $J=0$, $D=3/2$ and initial conditions $\theta_1(0)=0,
  \theta_2(0)=2\pi/3+\pi/4, \theta_3(0)=7\pi/6$; In (d), we use $G=1$, $J=0.1$,
  $D=2$ and initial conditions $\theta_1(0)=0, \theta_2(0)=0.1,
  \theta_3(0)=-0.2$. In all plots in Fig.~1, we fix $\gamma=P=3$ and
  $N_1(0)=6\times 10^{-6}$, $N_2(0)=3\times 10^{-6}$, $N_3(0)=4\times
  10^{-6}$.}\BibitemShut {Stop}%
\bibitem [{lim({\natexlab{a}})}]{limit_cycle_note1}%
  \BibitemOpen
  \href@noop {} {} \ \bibinfo {note} {Here, we identify
  $\psi_4\equiv \psi_1$ and $\psi_0\equiv \psi_3$. We also set $\hbar=1$ in the
  following dicussion.}\BibitemShut {Stop}%
\bibitem [{Fig()}]{Fig2parameter}%
  \BibitemOpen
  \href@noop {} {}\bibinfo {note} {All plots are obtained from numerical solutions of the generalized Josephson equations with initial conditions near the ferromagnetic fixed point. Here we use the initial conditions: $N_1(0)
  = 6 \times 10^{-6}$, $N_2(0) = 3 \times 10^{-6}$, $N_3(0) = 4 \times
  10^{-6}$, $\theta_{21}(0) = -0.04$, and $\theta_{32}(0) = 0.05$.}\BibitemShut
  {Stop}%
\bibitem [{cen()}]{centermass}%
  \BibitemOpen
  \href@noop {} {}\bibinfo {note} {We note that the generalized Josephson
  equations depend only on the relative phases of the three condensates. The
  center-of-mass phase is a zero mode and does not affect the
  dynamics.}\BibitemShut {Stop}%
\bibitem [{fig()}]{fig4parameter}%
  \BibitemOpen
  \href@noop {} {}\bibinfo {note} {We use initial conditions near the vortex
  fixed point: $\theta_{21}(0) = 2\pi/3 - 0.03$, $\theta_{32}(0) = 2\pi/3 +
  0.04$, $N_1(0) = 6 \times 10^{-6}$, $N_2(0) = 3 \times 10^{-6}$, and $N_3(0)
  = 4 \times 10^{-6}$.}\BibitemShut {Stop}%
\end{thebibliography}

%

\end{document}


\title{ Supplemental Material for \\``\textit{Emergent ac Effect  in Nonreciprocal Coupled   Condensates}"}

\author{Ji Zou}
\affiliation{Physics Department, King Fahd University of Petroleum and Minerals, 31261, Dhahran, Saudi Arabia}
\affiliation{Quantum Center, KFUPM, Dhahran, Saudi Arabia}
\author{Valerii K. Kozin}
\affiliation{Department of Physics, University of Basel, Klingelbergstrasse 82, 4056 Basel, Switzerland}
\author{Daniel Loss}
\affiliation{Physics Department, King Fahd University of Petroleum and Minerals, 31261, Dhahran, Saudi Arabia}
\affiliation{Quantum Center, KFUPM, Dhahran, Saudi Arabia}
\affiliation{Department of Physics, University of Basel, Klingelbergstrasse 82, 4056 Basel, Switzerland}
\affiliation{RDIA Chair in Quantum Computing}
\author{Jelena Klinovaja}
\affiliation{Department of Physics, University of Basel, Klingelbergstrasse 82, 4056 Basel, Switzerland}

\begin{abstract}
\end{abstract}
\maketitle

 \begin{figure}[!b]
	\centering\includegraphics[width=0.9\linewidth]{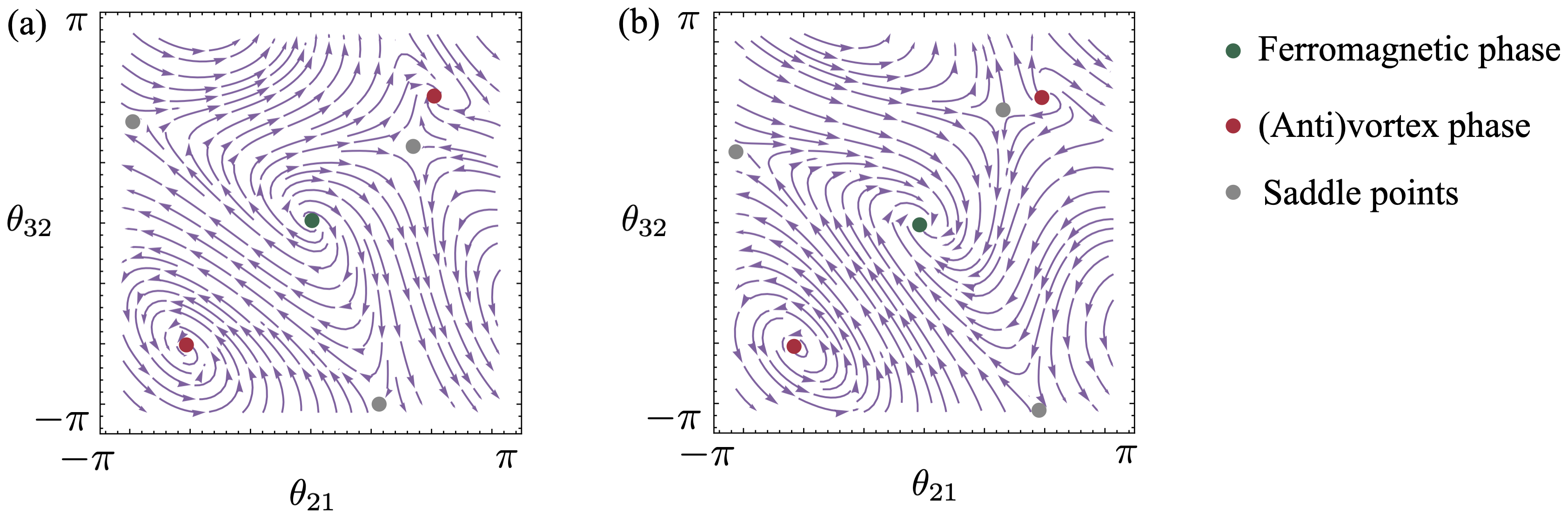}
 \caption{Phase space flow on a torus for the nonreciprocal Kuramoto model. (a) For $G = 1$ and $J = D = 2$, the ferromagnetic state is unstable, and the (anti)vortex state is stable. (b) For $G = -1$ and $J = D = 2$, the ferromagnetic state is stable, while the (anti)vortex state is unstable. In both cases, there are three saddle points on the torus. }
  \label{figsm1}
\end{figure}

\subsection{(i) Nonreciprocal Kuramoto model}
In this section, we provide a more detailed analysis of the nonreciprocal Kuramoto model. As discussed in the main text, in the steady-state limit the particle number becomes constant and identical on all three sites due to the rotation symmerty. In this case, the dynamical equations reduce to:
\(\al{ & \dot{\theta}_1=-\vec n(\theta_{12})\cdot \vec J_L-\vec n(\theta_{13})\cdot \vec J_R , \\
       & \dot{\theta}_2=-\vec n(\theta_{21})\cdot \vec J_R-\vec n(\theta_{23})\cdot \vec J_L ,\\
    & \dot{\theta}_3= -\vec n(\theta_{32}) \cdot \vec J_R -\vec n(\theta_{31}) \cdot \vec J_L.   } \label{s1}  \)
Here we have constant effective coupling vectors:
\( \vec J_L=(J, D-G/2), \;\;\;\text{and} \;\;\; \vec J_R=(J, -D-G/2),   \)
which describe the clockwise and counterclockwise particle hopping. 
We note that, in the special case $J=G=0$, the above model reduces to the Kuramoto model of synchronization. In general, we obtain a nonreciprocal Kuramoto model with nonreciprocal couplings between condensates. To study the possible steady states of this model, we observe that the dynamics is governed only by the relative phases. We therefore choose $\theta_{21}$ and $\theta_{32}$ as independent variables, whose dynamics is given by
\( \al{     \dot{\theta}_{21}& =  [\vec n(-\theta_{21}-\theta_{32}) -\vec n(\theta_{21})]\cdot \vec J_R  +[\vec n(-\theta_{21}) -\vec n(-\theta_{32})]\cdot \vec J_L , \\
   \dot{\theta}_{32} &=[ \vec n(\theta_{21}) - \vec n(\theta_{32})  ]\cdot \vec{J}_R  + [ \vec n(-\theta_{32}) - \vec n(\theta_{32}+\theta_{21})  ]\cdot \vec{J}_L.   }    \)
It is easy to check that  
\( (\theta_{21}, \theta_{32})=(0, 0)\;\; \rm{and} \ (\theta_{21}, \theta_{32})=(\pm 2\pi/3, \pm 2\pi/3),  \)  
are the fixed points of the dynamics such that $  \dot{\theta}_{21}=  \dot{\theta}_{32}=0$, corresponding to the ferromagnetic and (anti)vortex phase. To see if there are also other solutions to $  \dot{\theta}_{21}=  \dot{\theta}_{32}=0$, we note that they can be solved analytically when $J=0$. In this case, we find other three fixed points: $(\theta_{21}, \theta_{32})=(0, \pi), (\pi, 0), (\pi, \pi)$. However, unlike the ferromagnetic and (anti)vortex fixed points,  these three points will move around on the torus $(\theta_{21}, \theta_{32})\in \mathbb{T}^2$, when one changes the value of $J$. Importantly, these extra fixed points are always the saddle points and cannot be stable. This follows from the Poincaré–Hopf theorem: the sum of the indices of all singular points of a vector field on a two-dimensional compact orientable manifold must equal the Euler characteristic of the manifold. As we will see, the three  fixed points $(0, 0), (\pm 2\pi/3, \pm 2\pi/3)$ always have index +1. Since the Euler characteristic of the torus is zero, the remaining three fixed points must each have index -1, corresponding to saddle points. This can be also seen intuitively from numerics. As we show in Fig.~\ref{figsm1}, ferromagnetic and (anti)vortex fixed points are highlighted, which have an index of +1. One can also identify three saddle points in each plot.

 \begin{figure}[!t]
	\centering\includegraphics[width=0.88\linewidth]{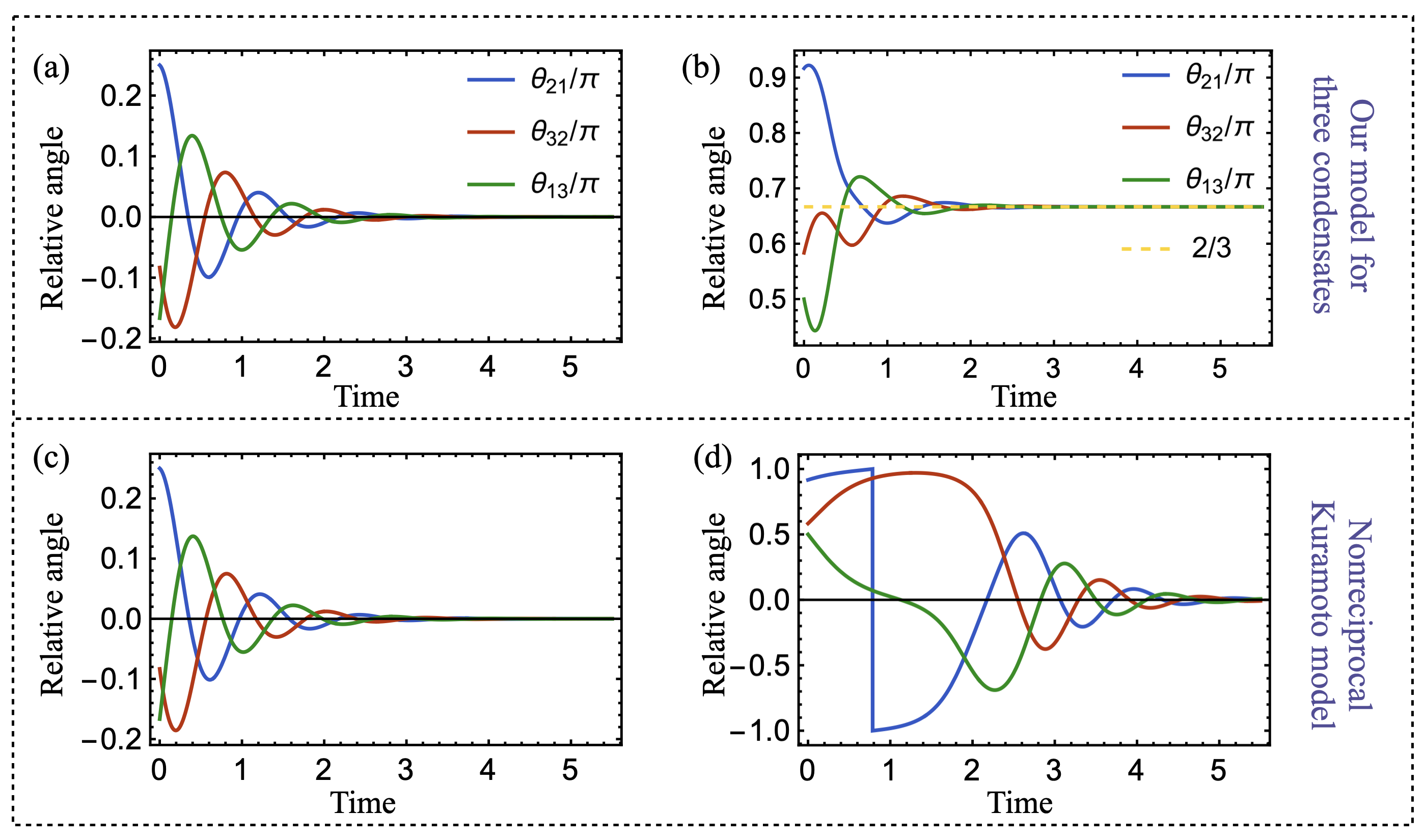}
\caption{
All panels use the same parameters: $\gamma = 3,\; G = -1,\; J = 0,\; D = 3, \; \gamma=3$. 
Panels (a) and (b) show the relative-angle dynamics of the three condensates governed by the generalized Josephson equations, {Eqs.~(3) and (4) in the main text}, starting from initial conditions near the ferromagnetic and vortex fixed points, respectively. In this regime, both fixed points are stable: the trajectory relaxes back to the ferromagnetic state in (a) and to the vortex state in (b). 
In contrast, in panels (c) and (d), where the angles evolve according to the nonreciprocal Kuramoto model { Eq.~\eqref{s1}} and the same two types of initial conditions are used, only the ferromagnetic state remains stable, as analyzed in the main text. In both (c) and (d), the trajectories converge to the ferromagnetic phase. {For all plots, we use initial conditions: $\theta_{1}(0)=0,\ \theta_{2}(0)={\pi}/{4},\ \theta_{3}(0)={\pi}/{6}$. For Panels (a) and (b), we use: $\ N_{1}(0)=6\times10^{-6},\ N_{2}(0)=3\times10^{-6},\ N_{3}(0)=4\times10^{-6}.$}
}

  \label{figsm11}
\end{figure}  

To determine the condition under which the ferromagnetic fixed point is stable, we introduce small perturbations around this fixed point. The generalized Josephson equations then reduce to:
\( \dv{}{t}\mqty(\delta \theta_{21}\\ \delta \theta_{32})=\mathcal{L}_F \mqty(\delta \theta_{21}\\ \delta \theta_{32}),    \) 
with the linear operator $\mathcal{L}_F$ being $\mathcal{L}_F={3G}/{2} + D(\hat\sigma_z+2i\hat\sigma_y)$. Here $\hat{\sigma}_{x,y,z}$ are the standard Pauli matrices. The stability of this fixed point is determined by the eigenvalues of this linear operator: $\lambda_{\pm}^F=3G/2\pm \sqrt{3}D i$. When the real parts of all the eigenvalues are negative, namely $G<0$, the ferromagnetic fixed point is stable [see Fig.~\ref{figsm1}(b)]. We also note that the real parts of the eigenvalues at this fixed point always share the same sign, and therefore its index is always +1.

One can perform a similar analysis for the (anti)vortex fixed point. For example, for the vortex fixed point, it is easy to show that the corresponding linear operator is given by:
\(\mathcal{L}_V = -\frac{3G}{4} - \frac{D-\sqrt{3}J}{2} (\hat{\sigma}_z+2i\hat{\sigma}_y),\)
which has eigenvalues 
\( \lambda_{\pm}^V= \frac{1}{4}[ -3G\pm i2 \sqrt{3} (D-\sqrt{3}J) ].  \)
Therefore, when $G>0$, the vortex phase is stable [see Fig.~\ref{figsm1}(a)].  We again note that the real parts of the eigenvalues at this fixed point always have the same sign, so its index is also +1.

Therefore, in the nonreciprocal Kuramoto model, the final stable state is always either the ferromagnetic phase or the (anti)vortex state, determined solely by the sign of the dissipative coupling $G$. However, in the general condensate dynamics governed by Eqs. (3) and (4) in the main text, the ferromagnetic phase and the (anti)vortex state can both be stable simultaneously for the same set of parameters, leading to hysteretic behavior. We show an example in Fig.~\ref{figsm11} for $G = -1,\; J = 0,\; D = 3, \;\gamma=3$.   In Figs.~\ref{figsm11}(a, b), depending on whether the initial condition is prepared near the ferromagnetic or vortex fixed point, the system relaxes into either attractor, confirming the coexistence of multiple stable phases. The stability conditions of the ferromagnetic and vortex phases are discussed in more detail below. In contrast, the nonreciprocal Kuramoto model in Figs.~\ref{figsm11}(c, d) admits only the ferromagnetic attractor for the same parameters.

\subsection{(ii) Instability of the ferromagnetic phase}
As we mentioned in the main text, the full phase space is six-dimensional, $(\theta_i, N_i) \in \mathbb{T}^3 \times \mathbb{R}_+^3$. Exploiting the global U(1) symmetry, we reduce it to a five-dimensional manifold $\mathcal{M}=\mathbb{T}^2 \times \mathbb{R}_+^3$, with coordinates given by two independent relative phases and the particle numbers: $(\theta_{21}, \theta_{32}, N_i)$. The dynamics of the system on $\mathcal{M}$ is governed by the generalized Josephson equations given in the  main text. In the ferromagnetic phase, all condensate phases are locked, corresponding to the fixed point $\vb*{\mathcal{V}}_F=(0, 0, N_i=N_0)$ with $N_0=\gamma/[2(\gamma+G)]$ in the phase space $\mathcal{M}$. {Here $N_0$ is determined by Eq. (4) in the main text by setting $dN_i/dt=0$ and $\theta_{21}=\theta_{32}=0$. } We wish to examine the condition under which the ferromagnetic state becomes unstable to arbitrary small fluctuations. To this end, we introduce small perturbations $\delta\vb*{\mathcal{V}}_F$ around the point $\vb*{\mathcal{V}}_F$. The generalized Josephson equations {Eqs.~(4) and (5) in the main text} then reduce to the following form:
\(  \partial_t \delta\vb*{\mathcal{V}}_F = \hat{\mathcal{L}}_F \delta\vb*{\mathcal{V}}_F, \;\;\text{where the linear operator takes the form}\;\;  \hat{\mathcal{L}}_F= \mqty(  \mathcal{L}_F^{\theta} &  J \mathcal{L}_1 \\ J\mathcal{L}_2  &  \mathcal{L}_F^{N} ).  \) 
Here, $\mathcal{L}_F^{\theta}$ and $\mathcal{L}_F^{N}$ are operators acting on $\mathbb{T}^2$ and $\mathbb{R}_+^3$, respectively, and are explicitly given by:
\( \mathcal{L}_F^{\theta}=\frac{3G}{2} + D(\hat\sigma_z+2i\hat\sigma_y)   , \;\;\;\text{and}\;\;\; \mathcal{L}_F^{N}= -(G+2\gamma) -\left(\frac{G}{2}-D\right) \hat{\mathcal{C}} -\left(\frac{G}{2}+D\right) \hat{\mathcal{C}}^{-1},    \)
where $\hat{\sigma}_{y,z}$ are the Pauli matrices and $\hat{\mathcal{C}}$ is the cyclic permutation matrix with elements $\mathcal{C}_{ij} = \delta_{j - i, 1}$~\footnote{Here, the difference $j - i$ is understood modulo 3.}. Here, $i, j\in \{1, 2, 3\}$. 
These two sectors are coupled to each other through the coherent symmetric coupling $J$, with 
\(  \mathcal{L}_1= \frac{3(G+\gamma)}{\gamma} \mqty( -1 & 1 & 0 \\ 0  & -1 & 1  ),  \;\;\;\text{and}\;\;\;   \mathcal{L}_2= \frac{\gamma}{G+\gamma} \mqty(2 & 1 \\ -1  & 1 \\ -1 & -2).  \)

As discussed in the main text under ``\textit{Model and generalized Josephson equations}" and in the section above, when the particle numbers $N_i$ are held fixed, reducing the phase space to $\mathbb{T}^2$, only two types of steady states are possible: the ferromagnetic and (anti)vortex phases, with their stability solely determined by the sign of $G$. The dynamical flows on $\mathbb{T}^2$ are shown in Fig.~\ref{figsm1}(a) for $G > 0$ and Fig.~\ref{figsm1}(b) for $G < 0$. For $G < 0$, the ferromagnetic state is stable and small perturbations decay. In contrast, for $G > 0$, the ferromagnetic state becomes unstable, and trajectories flow toward the (anti)vortex fixed points. This behavior is captured by the operator $\mathcal{L}_F^{\theta}$, which is the same as the linear operator  at the ferromagnetic fixed point for the nonreciprocal Kuramoto model with eigenvalues $\pm i \sqrt{3} D + 3G/2$. In sharp contrast, the operator $\mathcal{L}_F^N$ is always Hurwitz, with eigenvalues $-2(\gamma + G)$ and $-(G/2+2\gamma)\pm \sqrt{3}D i$, all of which have negative real parts when the system is stable $|G| < \gamma$. The stable number-sector dynamics can help stabilize the ferromagnetic phase via its coupling to the relative phase dynamics through $J$. 
The instability condition can be obtained analytically by evaluating the eigenvalues of the linear operator $\hat{\mathcal{L}}_F$. They are given by $\lambda_1 = -2(G+\gamma)$, $\lambda_{2,3} = \pm i\sqrt{3}D + G/2 - \gamma - \sqrt{(G+\gamma)^2 - 9J^2}$, and $\lambda_{4,5} = \pm i\sqrt{3}D + G/2 - \gamma + \sqrt{(G+\gamma)^2 - 9J^2}$. The real parts of $\lambda_{4,5}$ may become positive, signaling exponential growth of perturbations and hence instability of the ferromagnetic phase. The condition $\text{Re}(\lambda_{4,5}) < 0$ then yields the stability criterion given by Eq.~(8) in the main text, $G(G+4\gamma)<12J^2$. 

\begin{figure}[h]
	\centering\includegraphics[width=\linewidth]{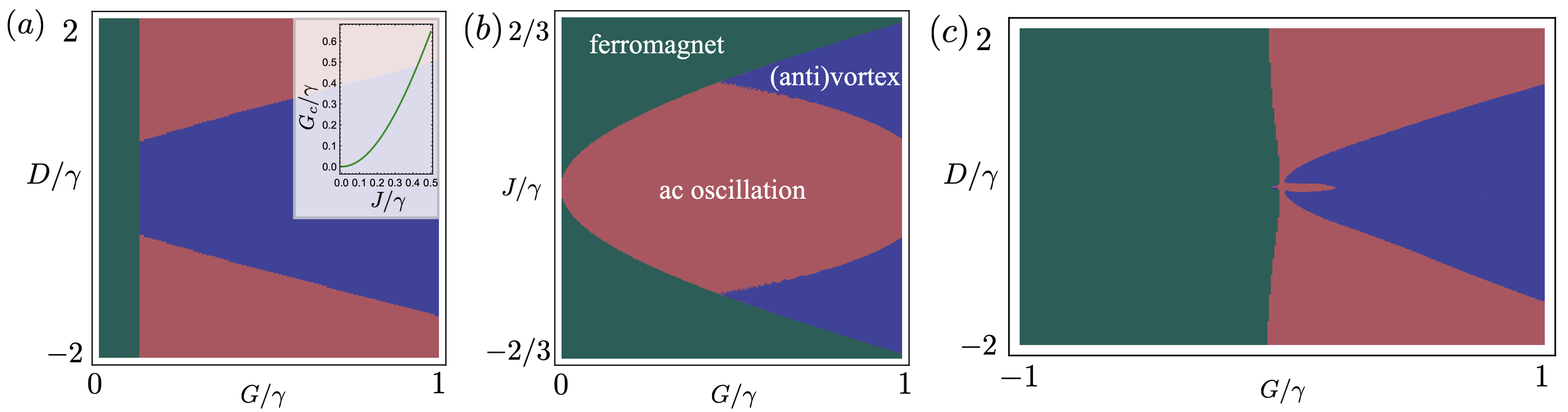}
 \caption{ (a) Phase diagram for finite $J$, with $J/\gamma = 0.2$. Inset: critical dissipative coupling $G_c$ as a function of the symmetric coherent coupling $J$. (b) Phase diagram in the $(G, J)$ plane at fixed $D/\gamma = 1.5$. (c) Phase diagram in the $(G, D)$ plane with coupling disorder: $D_{12} = 0.9D$, $D_{23} = 1.2D$, $D_{13} = D$; and $G_{12} = 0.9G$, $G_{23} = 0.7G$, $G_{13} = G$; and $J_{12} = 100 \,\text{kHz}$, $J_{23} = -200\, \text{kHz}$, $J_{13} = 0$ with $\gamma=3\, \text{MHz}$. Here $J_{ij}, D_{ij}, G_{ij}$ stand for the couplings between the condensates $i$ and $j$. All plots are obtained by  numerically solving the generalized Josephson equations starting from initial conditions near the ferromagnetic fixed point.}
  \label{figsm2}
\end{figure}

We note that the ferromagnetic phase remains stable for all $G < 0$. When $J = 0$, it becomes unstable as $G$ turns positive—the same as  the case where particle numbers $N_i$ are held fixed. However, the final state differs: in the full dynamics, an intermediate ac oscillatory phase emerges before reaching the (anti)vortex regime, as discussed in the main text. Figure~\ref{figsm2}(a) shows the resulting phase diagram for finite $J$ for positive $G$. As expected, finite coherent coupling stabilizes the ferromagnetic phase over a window of positive $G$, with a critical boundary given analytically by $G_c = 2\sqrt{3J^2 + \gamma^2} - 2\gamma$ [inset in Figure~\ref{figsm2}(a)]. While the ferromagnetic instability threshold can be obtained analytically, the transition from the ac phase to the (anti)vortex state lies deep in the nonlinear regime and is determined numerically. In Fig.~\ref{figsm2}(b), we also plot the phase diagram as a function of $J$ and $G$ for fixed $D$. The ferromagnetic boundary, $G(G+4\gamma) = 12J^2$, is independent of $D$, while the size of the ac region increases with $|D|$. 

To ensure the experimental viability of the ac phase, we examine its robustness against disorder by introducing variations in the coupling strengths between condensates. As shown in Fig.~\ref{figsm2}(c), the ac phase persists and is robust against the small variations, though the phase boundaries become distorted.

\subsection{(iii) Instability of the vortex phase}
\begin{figure}[h]
	\centering\includegraphics[width=\linewidth]{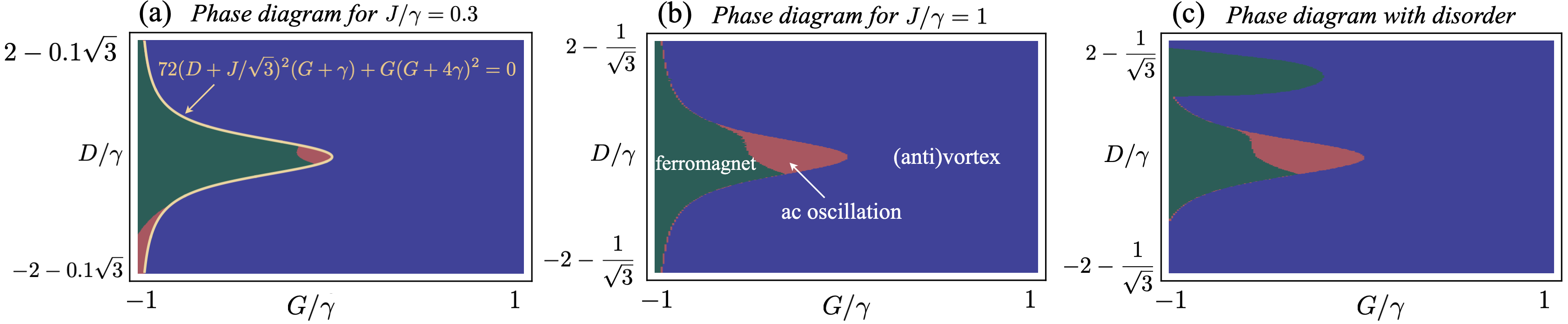}
 \caption{ (a) Phase diagram as a function of the antisymmetric coupling $D$ and dissipative coupling $G$ for finite $J$, with $J/\gamma = 0.3$. The light yellow curve is the instability condition obtained analytically, which agrees well with the numerical result.   (b) Phase diagram with $J/\gamma = 1$. (c) Phase diagram with coupling disorder: $D_{12} = 0.9D$, $D_{23} = 1.1D$, $D_{13} = D$; and $G_{12} = 0.8G$, $G_{23} = 0.9G$, $G_{13} = G$; and $J_{12} =1.1 J$, $J_{23} =0.9 J$, $J_{13} = J$ with $J/\gamma=1$. Here $J_{ij}, D_{ij}, G_{ij}$ stand for the couplings between the condensates $i$ and $j$.  All plots are obtained by  numerically solving the generalized Josephson equations starting from initial conditions near the vortex fixed point.}
  \label{figsm3}
\end{figure}

Here we expand the analysis of the stability of the vortex phase. As in the previous case, this is done by introducing small perturbations $\delta \vb*{\mathcal{V}}_V$ around the vortex fixed point $\vb*{\mathcal{V}}_V = (2\pi/3, 2\pi/3, \tilde{N}_0)$, with $\tilde{N}_0 = \gamma / (2\gamma - G)$. {Here $\tilde{N}_0$ is determined by Eq. (4) in the main text by setting $dN_i/dt=0$ and $\theta_{21}=\theta_{32}=2\pi/3$. } The generalized Josephson equations then reduce to:
\(   \partial_t \delta\vb*{\mathcal{V}}_V = \hat{\mathcal{L}}_V \delta\vb*{\mathcal{V}}_V, \;\;\text{where the linear operator takes the form}\;\;  \hat{\mathcal{L}}_V= \mqty(  \mathcal{L}_V^{\theta} &   \tilde{\mathcal{L}}_1 \\  \tilde{\mathcal{L}}_2  &  \mathcal{L}_V^{N} ).  \) 
The operators $\mathcal{L}_V^{\theta}$ and $\mathcal{L}_V^{N}$ acting on the subspaces $\mathbb{T}^2$ and $\mathbb{R}_+^3$ are given by: 
\( \mathcal{L}_V^{\theta} = -\frac{3G}{4} - \frac{D-\sqrt{3}J}{2} (\hat{\sigma}_z+2i\hat{\sigma}_y),\;\;\; \text{and}\;\;\;  \mathcal{L}_V^{N} =-2\gamma+\frac{G}{2} - \frac{1}{2} \left( D-\sqrt{3} J -\frac{G}{2} \right) \hat{\mathcal{C}} + \frac{1}{2} \left( D-\sqrt{3}J+\frac{G}{2} \right) \hat{\mathcal{C}}^{-1}.  \)
It is clear that they take a different form compared to the ferromagnetic case. Here, $\mathcal{L}_V^{\theta}$ is the same as the linear operator at the vortex fixed point in the nonreciprocal Kuramoto model. 
We note that $\mathcal{L}_V^{N}$ is always Hurwitz. The coupling between the phase sector $\mathbb{T}^2$ and the number sector $\mathbb{R}_+^3$ is encoded in the off-diagonal blocks $\tilde{\mathcal{L}}_{1,2}$. Specifically, $\tilde{\mathcal{L}}_1$ captures how fluctuations in $\mathbb{R}_+^3$ drive the dynamics of relative phases in $\mathbb{T}^2$:
\(\tilde{\mathcal{L}}_1 = \frac{3(J+\sqrt{3}D)}{4\gamma/(G-2\gamma)}\mqty(-1 & 1 & 0 \\ 0 & -1 & 1 ) + \frac{\sqrt{3}G}{8\gamma/(G-2\gamma)} \mqty(-1 & -1 & 2 \\ 2 & -1 & -1)   \)
   On the other hand, $\tilde{\mathcal{L}}_2$ describes how the relative phase dynamics in $\mathbb{T}^2$ affect the particle number evolution in $\mathbb{R}_+^3$:
\(  \tilde{\mathcal{L}}_2 = \frac{J+\sqrt{3}D}{(G-2\gamma)/\gamma}\mqty(2 & 1\\ -1 & 1 \\ -1 & -2) + \frac{\sqrt{3}G}{2(G-2\gamma)/\gamma}\mqty( 0 & 1\\ -1 & -1 \\1 &0 ).     \)
These coupling matrices also differ qualitatively from those in the ferromagnetic case. Notably, the antisymmetric exchange $D$ enters $\tilde{\mathcal{L}}_{1,2}$ in the vortex case but is absent in the ferromagnetic stability operator $\mathcal{L}_{1,2}$. The stability of the vortex phase can be analyzed by evaluating the eigenvalues of the linear operator $\hat{\mathcal{L}}_V$. Although more involved, the eigenvalues can still be obtained analytically. They include a real mode $-2\gamma + G$ and two complex-conjugate pairs:
\(\al{ &\lambda_{\pm}=\frac{1}{4}\left[  -2i\sqrt{3}(D-\sqrt{3}J) -G-4\gamma \pm \sqrt{-108 ( D+J/\sqrt{3} )^2 -36i\sqrt{3}G(D+J/\sqrt{3}) +13G^2 -16G\gamma +16\gamma^2 }   \right],  \\ 
   &\tilde{\lambda}_{\pm}=\frac{1}{4}\left[  2i\sqrt{3}(D-\sqrt{3}J) -G-4\gamma \pm \sqrt{-108 ( D+J/\sqrt{3} )^2 +36i\sqrt{3}G(D+J/\sqrt{3}) +13G^2 -16G\gamma +16\gamma^2 }   \right].  } \)
The vortex phase becomes unstable when any of the real parts of these eigenvalues becomes positive. This leads to the analytical instability condition: $72(D+J/\sqrt{3})^2 (G+\gamma)+G(G+4\gamma)^2<0$, which is the equation (9) in the main text. This condition accurately traces the numerically determined vortex phase instability boundary, as depicted in Fig.~\ref{figsm3}(a).  There, the  yellow curve stands for the analytical instability condition, and is plotted on the phase diagram that is obtained numerically.

We show the phase diagram for the $J = 0$ case in the main text. Figures~\ref{figsm3}(a) and (b) present the corresponding diagrams for finite exchange couplings, $J/\gamma = 0.3$ and $J/\gamma = 1$. Two key effects of $J$ are evident. First, it shifts the vortex instability boundary along the $D$ axis by $-J/\sqrt{3}$ (note we also shift the range of $D$ in the plots). Second, while the ac phase shrinks in the region around $G/\gamma \sim -1$ as $J$ increases, a new window of ac oscillation phase appears near $G/\gamma \lesssim 0$. Thus, the ac phase persists across a broad range of parameters. In Fig.~\ref{figsm3}(c), we further demonstrate the robustness of the ac phase in the presence of disorder introduced via small variations in coupling strengths. These results suggest that the ac phase is robust and does not rely on fine-tuning, making it amenable to experimental observation.

\subsection{(iv) Hysteresis in the dynamics}
\begin{figure}[h]
	\centering\includegraphics[width=0.78\linewidth]{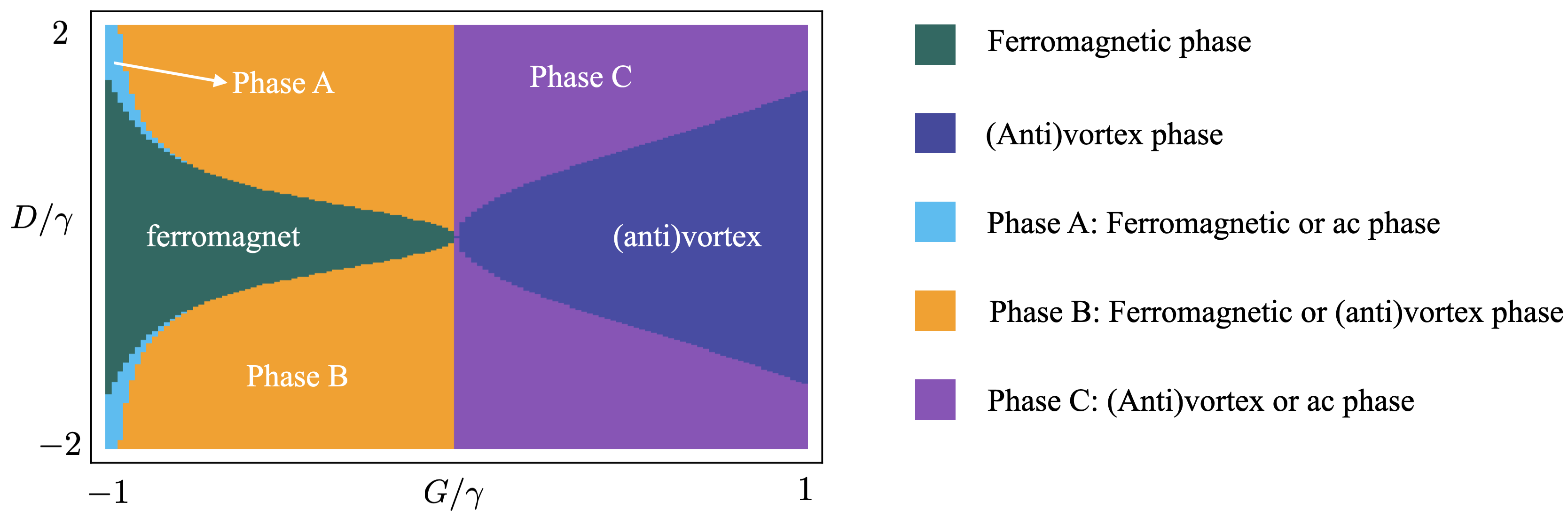}
 \caption{ {Phase diagram as a function of $G/\gamma$ and $D/\gamma$ at 
$J=0$. Colors indicate the nature of the long time steady state. There are regimes where the system always relaxes to a unique ferromagnetic or (anti)vortex state, respectively, independent of the initial condition. Phase A corresponds to coexistence of ferromagnetic and ac phases, Phase B to coexistence of ferromagnetic and (anti)vortex phases, and Phase C to coexistence of (anti)vortex and ac phases. In these multistable regimes, the final state depends on the initial condition or history, giving rise to hysteresis in the system dynamics. The plot is obtained by numerically solving the generalized Josephson equations [Eqs. (3) and (4) in the main text] for different initial conditions, chosen close to the ferromagnetic and vortex states.}}
  \label{figsm4}
\end{figure}

{We emphasize that the instability condition of vortex phase we discussed above [Eq. (9) in the main text] is qualitatively different from that of the ferromagnetic state [Eq.~(8) in the main text and Fig.~\ref{figsm2}]. This is expected, since the analysis is performed in the vicinity of distinct fixed points that carry different topological structure. It is evident that, in certain regions of parameter space, multiple steady states coexist, including ferromagnetic, (anti)vortex, and ac phases. This multistability underlies the hysteretic behavior discussed in the main text. In Fig.~\ref{figsm4}, we present the phase diagram for $J=0$, obtained by considering different initial conditions, chosen near the ferromagnetic and (anti)vortex states. As shown in Fig.~\ref{figsm4}, there exist parameter regimes in which the long time state is uniquely determined, independent of the initial condition. In these regions, the system always relaxes either to a ferromagnetic state or to an (anti)vortex state. In contrast, other regimes display a pronounced dependence on the initial condition or, equivalently, on the history of the system. In these regions, distinct dynamical phases can be stabilized for identical parameters [see Fig.~\ref{figsm4}]: Phase A corresponds to parameters for which the final state can be either ferromagnetic or an ac phase. Phase B denotes regimes where the system relaxes either to a ferromagnetic or an (anti)vortex state. Phase C represents regimes in which the final state can be either an (anti)vortex or an ac phase. The existence of such bistable regions directly leads to hysteresis in the dynamics of the system.   }

\subsection{(v) Amplitude of the oscillations of relative angles}
In the main text we state that at the Hopf threshold, oscillations of relative angles emerge with the amplitude $A$, whose square is proportional to the real part of the corresponding eigenvalue $A^2\sim G/2-\gamma+\sqrt{(G+\gamma)^2-9J^2}$. Below we outline, following Kuzntesov's textbook~\cite{Kuznetsov}, how the missing coefficient of proportionality can be found. If we start in the ferromagnetic phase and go beyond the linear order (the generalized Josephson equations in Section (ii) in SM), we may write the resulting system of equations as 
\(   \partial_t \delta\vb*{\mathcal{V}}_F = \hat{\mathcal{L}}_F \delta\vb*{\mathcal{V}_F} + \vb*{\mathcal{F}}(\delta\vb*{\mathcal{V}}_F),\)
where $\vb*{\mathcal{F}}(\delta\vb*{\mathcal{V}}_F)$ represents non-linear corrections, that we have to include up to the third order in $\delta\vb*{\mathcal{V}}_F$. At the Hopf threshold, $\hat{\mathcal{L}}_F$ has a simple pair of complex eigenvalues on the imaginary axis, and these are the only eigenvalues with zero real part (in our case right at the threshold they are given by $\lambda=\pm i \sqrt{3} D$). Let's denote the corresponding right and left eigenvectors of $\hat{\mathcal{L}}_F$ as $\hat{\mathcal{L}}_F \vb*{q}=\lambda \vb*{q}$ and $\hat{\mathcal{L}}_F^T \vb*{p}=-\lambda \vb*{p}$. The left and right eigenvectors are normalized as $\langle \vb*{p},\vb*{q}\rangle=1$, where the scalar product is defined as $\langle \vb*{x},\vb*{y} \rangle=\sum_i x^*_i y_i$. Then, defining the following matrix-valued functions 
\({\mathcal{B}_i}(\vb*{x},\vb*{y})=\sum_{j,k=1}^{5} \frac{\partial^2 \mathcal{F}_i(\vb*{\xi})}{\partial \xi_j \partial \xi_k} \bigg|_{\vb*{\xi}=0}x_j y_k, \)
and
\({\mathcal{C}_i}(\vb*{x},\vb*{y},\vb*{z})=\sum_{j,k,l=1}^{5} \frac{\partial^3 \mathcal{F}_i(\vb*{\xi})}{\partial \xi_j \partial \xi_k \partial \xi_l} \bigg|_{\vb*{\xi}=0}x_j y_k z_l, \)
the First Lyapunov coefficient~\cite{Kuznetsov} can be expressed as
\( l_1= \frac{1}{2|\lambda |}\text{Re}\left[\langle \vb*{p},\vb*{\mathcal{C}}(\vb*{q},\vb*{q},\vb*{q}^*)\rangle-2\langle \vb*{p},\vb*{\mathcal{B}}(\vb*{q},\hat{\mathcal{L}}_F^{-1}\vb*{\mathcal{B}}(\vb*{q},\vb*{q}^*))\rangle+\langle \vb*{p},\vb*{\mathcal{B}}(\vb*{q}^*,(2i|\lambda|-\hat{\mathcal{L}}_F)^{-1}\vb*{\mathcal{B}}(\vb*{q},\vb*{q}))\rangle\right].\)	
The inverse of the  First Lyapunov coefficient $1/l_1$, multiplied by the corresponding component of the right eigenvector $\vb*{q}$ (due to the reduction of the system to its normal form by mapping the vector $\delta\vb*{\mathcal{V}}_F=z \vb*{q}+z^* \vb*{q}^* +\vb*{y}$ onto the complex variable $z=\langle\vb*{p},\delta\vb*{\mathcal{V}}_F\rangle\in \mathbb{C}^1$ with $\vb*{y}\in\mathbb{R}^5$) approximately yields the desired coefficient of proportionality of the amplitude ($A^2$) of the oscillations of relative angles after the bifurcation. We note, that expression for the first Lyapunov coefficient involves inversion of 5 by 5 matrices which makes obtaining exact analytical expressions cumbersome.


%